# Integration of Atomic Layer Epitaxy Crystalline $Ga_2O_3$ on Diamond for Thermal Management

Zhe Cheng[1], Virginia D. Wheeler[2], Tingyu Bai[3], Jingjing Shi[1], Marko J. Tadjer[2], Tatyana Feygelson[2], Karl D. Hobart[2], Mark S. Goorsky[3], Samuel Graham[1, 4, *]

[1] George W. Woodruff School of Mechanical Engineering, Georgia Institute of Technology, Atlanta, Georgia 30332, USA

[2] U.S. Naval Research Laboratory, 4555 Overlook Ave SW, Washington, DC 20375, USA

[3] Materials Science and Engineering, University of California, Los Angeles, Los Angeles, CA, 91355, USA

[4] School of Materials Science and Engineering, Georgia Institute of Technology, Atlanta, Georgia 30332, USA

[*] Corresponding author: sgraham@gatech.edu

# ABSTRACT

Gallium oxide ($Ga_2O_3$) has attracted great attention for electronic device applications due to its ultra-wide bandgap, high breakdown electric field, and large-area affordable substrates grown from the melt. However, its thermal conductivity is significantly lower than that of other wide bandgap semiconductors such as SiC, AlN, and GaN, which will impact its ability to be used in high power density applications. Thermal management in $Ga_2O_3$ electronics will be the key for device reliability, especially for high power and high frequency devices. Similar to the method of cooling GaN-based high electron mobility transistors by integrating it with high thermal conductivity diamond substrates, this work studies the possibility of heterogeneous integration of $Ga_2O_3$ with diamond for thermal management of $Ga_2O_3$ devices. In this work, $Ga_2O_3$ was deposited onto single crystal diamond substrates by atomic layer deposition (ALD) and the thermal properties of ALD-$Ga_2O_3$ thin films and $Ga_2O_3$-diamond interfaces with different interface pretreatments were measured by Time-domain Thermoreflectance (TDTR). We observed very low thermal conductivity of these $Ga_2O_3$ thin films (about 1.5 W/m-K) due to the extensive phonon grain boundary scattering resulting from the nanocrystalline nature of the $Ga_2O_3$ film. However, the measured thermal boundary conductance (TBC) of the $Ga_2O_3$-diamond interfaces are about 10 times larger than that of the Van der Waals bonded $Ga_2O_3$-diamond interfaces, which indicates the significant impact of interface bonding on TBC. Furthermore, the TBC of the Ga-rich and O-rich $Ga_2O_3$-diamond interfaces are about 20% smaller than that of the clean interface, indicating interface chemistry affects interfacial thermal transport. Overall, this study shows that a high TBC can be obtained from strong interfacial bonds across $Ga_2O_3$-diamond interfaces, providing a promising route to improving the heat dissipation from $Ga_2O_3$ devices with lateral architectures.

**INTRODUCTION**

As an emerging semiconductor material with an ultra-wide bandgap (4.8 eV) and high critical electric field (8 MV/cm), β-$Ga_2O_3$ has attracted great attention for electronic device applications.[1] However, the thermal conductivity of bulk β-$Ga_2O_3$ (10-30 W/m-K, depending on crystal orientations) is at least one order of magnitude lower than those of other wide bandgap semiconductors such as GaN (230 W/m-K), 4H-SiC (490 W/m-K), AlN (320 W/m-K), and diamond (>2000 W/m-K).[2-4] Size effects, doping, and alloying can further reduce the thermal conductivity of $Ga_2O_3$-based materials which is an essential thermophysical property that impacts device behavior and reliability.[4] As seen in other wide bandgap devices, higher device operating temperatures (as a result of the low thermal conductivity in β-$Ga_2O_3$) will result in faster device degradation and shorter lifetimes. Therefore, efficient thermal dissipation while minimizing device junction temperature is one of the main challenges for β-$Ga_2O_3$ -based devices, especially for high power and high frequency devices where localized Joule-heating results in hot-spots that can lead to degradation and failure.[5] While thermal issues in the use of β-$Ga_2O_3$ is a major concern, the majority of the characterization of $Ga_2O_3$ devices has been electrical in nature and only a few studies on thermal performance and thermal management have been reported.[6-9] Recently, exfoliated $Ga_2O_3$ nanomembrane field effect transistors (FET) have been fabricated on a single crystal diamond.[9] Even though the weak Van der Waals bonding of the $Ga_2O_3$ nanomembrane and the diamond substrate leads to a small thermal boundary conductance (TBC) (17 MW/m$^2$K), DC power density in excess of 60 W/mm demonstrated by β-$Ga_2O_3$ has been achieved on a similarly-exfoliated device.[5,9] This result demonstrates the great potential of $Ga_2O_3$ devices when integrated with a high thermal conductivity substrate, even if this integration approach (mechanical exfoliation and transfer of β-$Ga_2O_3$) is not readily scalable as a thermal management strategy for β-Ga$_2$O$_3$ devices. Thus, a scalable approach for integrating β-Ga$_2$O$_3$ with high thermal conductivity diamond is still lacking, which is essential for real-world applications.

In this work, we demonstrate growth of crystalline Ga$_2$O$_3$ on diamond, and investigate the quality of the resulting films as well as TBC between β-Ga$_2$O$_3$ and diamond substrates. Ga$_2$O$_3$ was deposited on single crystal diamond using atomic layer deposition (ALD) with three different diamond surface pretreatments as well as a reference substrate without pretreatment. The thermal properties of the ALD Ga$_2$O$_3$ thin films and the Ga$_2$O$_3$–diamond interfaces were measured using time-domain thermoreflectance (TDTR). Additionally, material characterization involving transmission electron microscopy (TEM) analysis was performed to understand the structure-thermal property relationship.

**SAMPLES AND METHODS**

Unlike the highly reactive plasma environment of chemical vapor deposited (CVD) diamond growth, ALD growth relies on the sequential self-limiting reactions between surface adsorbed metal organic molecules and oxidizing molecules. Single crystalline (100) diamond substrates were obtained commercially (Element Six, thermal grade) and cleaned using a sequence of treatments intended to remove metal and non-diamond carbon contamination: HNO$_3$:HCl, HNO$_3$:H$_2$SO$_4$, ultrasonic clean in ethanol, and finally an HF etch. The substrates were stored in ethanol and were dried in N$_2$ immediately prior to transfer into the Ga$_2$O$_3$ growth reactor. Thin films (~30-115 nm) were deposited on single crystal (100) diamond substrates in a Fiji 200 G2 reactor at 350°C using alternating cycles of trimethylgallium (TMG, STREM PURATREM) as the Ga precursor and a remote pure oxygen plasma as the oxidizing source. All samples utilized a turbo pump to drop the pressure in the chamber to 8 mTorr during plasma exposure. Under these conditions, the growth rate was 0.65Å/cycle.

To measure the thermal conductivity of the $Ga_2O_3$ film, a thick sample (about 120 nm $Ga_2O_3$) was grown on a single crystal diamond substrate (Samp1). Since this layer is too thick for our thermal measurement system to be sensitive to the interface TBC, thinner (~30nm) layers of $Ga_2O_3$ were grown onto other diamond substrates. In addition, the diamond substrates were given different in-situ surface pretreatments prior to growth to investigate the effect on TBC. The surface of Samp2 was pretreated with a Ga flashoff process to emulate the one typically used to clean surfaces in MBE. This consists of dosing with TMG to create a gallium sub-oxide at the surface that is subsequently exposed and removed with a hydrogen plasma pulse. The surface of Samp3 was initiated by super saturating with 10 consecutive Ga pulses prior to growth, while the surface of Samp4 received 10, 10s $O_2$ plasma pulses in-situ treatment prior to growth. The details of each sample is summarized in Table 1. A layer of Al (~84 nm) was deposited on the sample surface as a transducer for TDTR measurements. The thickness of the Al layer was determined by the picosecond acoustic technique during TDTR measurements. The thickness of $Ga_2O_3$ films are determined with TEM. TDTR is a pump-probe technique to measure thermal properties of both nanostructured and bulk materials.[10-12] A modulated pump beam (400 nm) heats the sample surface while a delayed probe beam (800 nm) detects the temperature variation of the sample surface. After being picked up by a photodetector and a lock-in amplifier, the signal is fitted with an analytical heat transfer solution to infer the unknown parameters.[5,13-15] More details about TDTR measurements on similar sample structures can be found in literature.[5,14,16]

## RESULTS AND DISCUSSION

Table 1. Sample structures and thermal properties

|  | **Al** | **$Ga_2O_3$** | **Diamond surface pretreatment** | **$Ga_2O_3$-C TBC** | **$Ga_2O_3$ $k$** |
|---|---|---|---|---|---|
| Samp1 | 88 nm | 115 nm | N/A | N/A | 1.76 W/m-K |
| Samp2 | 84 nm | 29 nm | Ultra-clean | 179 MW/m2-K | 1.50 W/m-K |
| Samp3 | 84 nm | 30 nm | Ga-rich | 136 MW/m2-K | 1.50 W/m-K |
| Samp4 | 84 nm | 28 nm | O-rich | 139 MW/m2-K | 1.52 W/m-K |

As shown in Table 1, the measured thermal conductivity of Samp1 is 1.76 W/m-K, which is much lower than that of bulk β-$Ga_2O_3$ (10-30 W/m-K).[4] To understand this low thermal conductivity value, TEM was used to study the fine structure, as shown in Fig. 1(a). The TEM image of Samp1 shows the $Ga_2O_3$ film is polycrystalline with grains on the order of 10-20nm. The yellow boundaries outline the grains and the red lines show the orientation of a few of the grains. The strong size effect resulting from small grains limits the phonon mean free path and correspondingly reduces the thermal conductivity of the $Ga_2O_3$ thin film. Furthermore, some areas are randomly ordered with rough grain boundaries which result in low phonon transmission and further decreases the thermal conductivity of the $Ga_2O_3$ thin film. The clean interface of $Ga_2O_3$ and diamond is shown in Fig. 1(b).

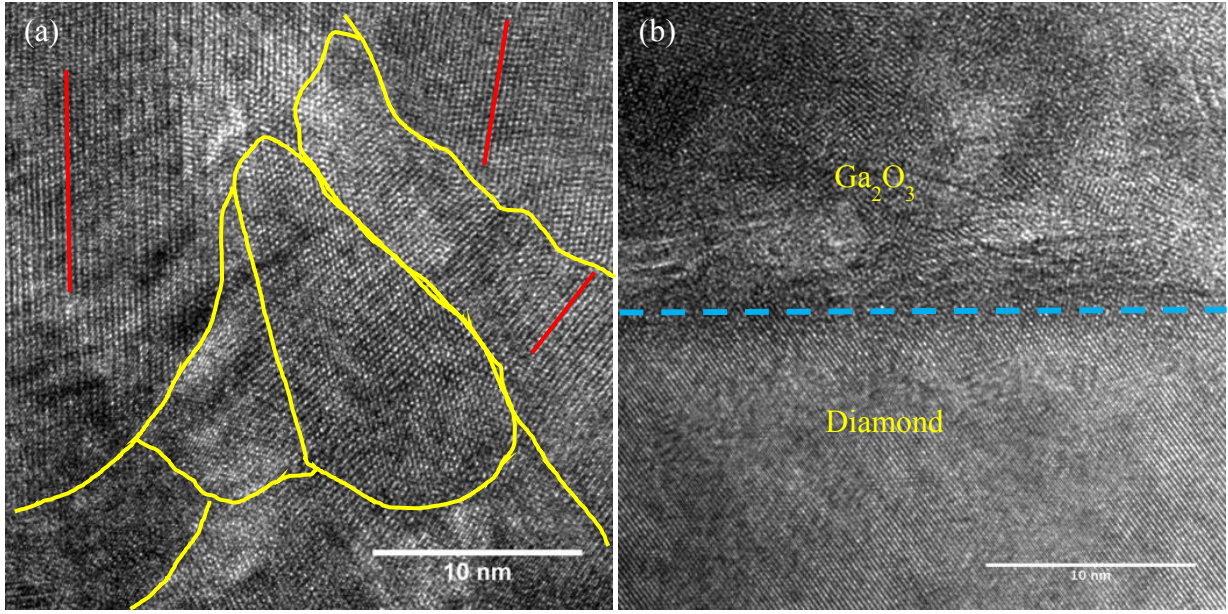

FIG. 1. (a) TEM image of Ga$_2$O$_3$ grain structure of Samp1 and (b) TEM image of Ga$_2$O$_3$-diamond interface of Samp1.

The measured thermal conductivity of Samp2-4 is about 1.5 W/m-K, which is much lower than the bulk counterpart and slightly lower than that of Samp1. By taking the small grain size (10-20 nm) into consideration, the film thickness (~30 nm) shows a slight thickness dependent thermal conductivity effect, but this effect plays a less important role in limiting phonon mean free paths. As shown in Fig. 2, the measured thermal conductivity of Ga$_2$O$_3$ thin films is compared with minimum thermal conductivity of amorphous Ga$_2$O$_3$ estimated by two minimum thermal conductivity models: the Cahill model and the diffuson model. The Cahill model is known as the amorphous limit of a material which assumes the mean-free-paths of the Debye-like, heat-carrying oscillations as half of the corresponding wavelengths. The corresponding minimum thermal conductivity is:[13,17]

$$\kappa_{Cahill} = \left(\frac{\pi}{6}\right)^{1/3} k_B n^{2/3} \sum_{i=1}^{3} v_i \left(\frac{T}{\Theta_i}\right)^2 \int_0^{\Theta_i/T} \frac{x^3 e^x}{(e^x-1)^2} dx \ , \qquad (1)$$

where $k_B$ is the Boltzmann constant, $v_i$ is the sound velocity of polarization $i$, $n$ is the atomic density, and $\Theta_i = v_i(h/2\pi k_B)(6\pi^2 n)^{1/3}$ is the cutoff frequency expressed as temperature unit where $h$ is the Planck constant.[13] The diffuson model of minimum thermal conductivity introduces a diffuson diffusivity and the corresponding minimum thermal conductivity is:[13,18]

$$\kappa_{diffuson} = \frac{n^{1/3}k_B}{\pi} \int_0^\infty \left(\frac{g(\omega)}{3n}\right)\left(\frac{h\omega}{2\pi T k_B}\right)^2 \frac{e^{\frac{h\omega}{2\pi T k_B}}}{\left(e^{\frac{h\omega}{2\pi T k_B}}-1\right)^2} \omega d\omega . \tag{2}$$

The only input in this model is the density of states $g(\omega)$, which is obtained from DFT calculations. We use the single crystal phase to approximate the density of states of amorphous phase because they are usually similar.[19]

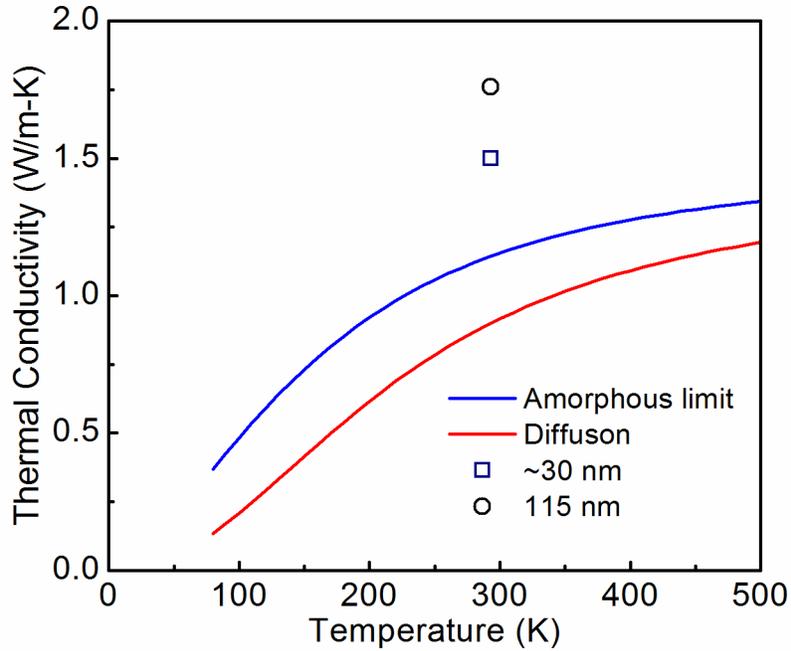

FIG. 2. the comparison of the measured thermal conductivity of $Ga_2O_3$ thin films with calculated minimum thermal conductivity. "Amorphous limit" is calculated based on Cahill mode of minimum thermal conductivity while "diffuson" is based on a diffuson model of minimum thermal conductivity.[13,17,18]

The calculated minimum thermal conductivity of $Ga_2O_3$ is 1.15 W/m-K and 0.90 W/m-K at room temperature for the amorphous limit and the diffuson model, respectively. Comparing to the bulk thermal conductivity (10-30 W/m-K), the measured thermal conductivity of the 30-nm-$Ga_2O_3$ thin films are only 1.3 times as the amorphous limit, and 1.67 times as the diffuson minimum thermal conductivity. Crystalline materials usually have a much larger thermal conductivity than their amorphous counterpart. Here, the fine nanocrystalline nature makes the thermal conductivity of the measured films so close to the thermal conductivity of amorphous $Ga_2O_3$. Figure 3(a-b) show the TEM images of $Ga_2O_3$ grown on diamond for Samp2 and Samp4, respectively. All these four samples (Samp1-4) have similar grain structures for these $Ga_2O_3$ thin films. There are some spotty contrast areas which are about 10 nm away from the $Ga_2O_3$-diamond interfaces for the thinner films in Fig. 3. These areas are possible amorphous or less crystalline regions which could be further affected or damaged by the focus ion beam (FIB) process during TEM sample preparation, which may contribute to the reduction in thermal conductivity of $Ga_2O_3$ thin films. The low thermal conductivity of ALD $Ga_2O_3$ also highlights the importance of the integration with diamond to extract heat from $Ga_2O_3$.

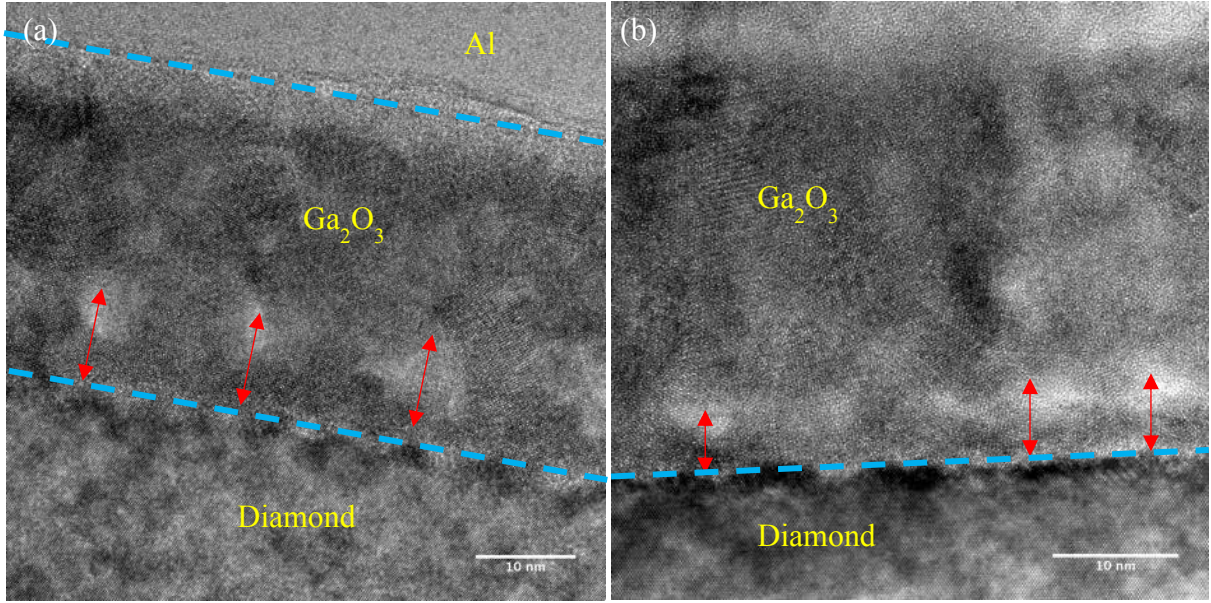

FIG. 3. TEM images of $Ga_2O_3$ grown on diamond: (a) Samp2 and (b) Samp4.

Higher-resolution TEM images of the $Ga_2O_3$-diamond interfaces are shown in Fig. 4, with Fig. 4(a) and Fig. 4(b) are for Samp2 and Samp4, respectively. The interfaces are not atomic smooth but are atomically abrupt between $Ga_2O_3$ and diamond without voids or exfoliation at the interfaces. This good contact explains the relatively high TBC of the $Ga_2O_3$-diamond interfaces reported in Table I above. The measured TBC of the ultra-clean (Samp2) interface is 179 MW/m$^2$-K, about 10 times higher than TBC of a Van der Waals bonded $Ga_2O_3$-diamond interface, suggesting that covalent bonding facilitates interfacial heat transport better than Van der Waals interfacial bonding.[5] While the other two thinner samples have smaller TBC due to different in-situ pretreatments of the diamond surfaces, they are still much larger than the Van der Waals bonded TBC. This confirms that the type of interface bonding affects TBC significantly, similar to metal-quartz interfaces reported before.[20] Like epitaxy, room-temperature surface activated bonding (SAB) technique bonds independently grown layers with covalent chemical bonding interfaces, resulting in high TBC.[16] Thus, from the results shown here, we expect SAB Ga$_2$O$_3$-related interfaces should have high TBC, enabling another possible approach for integrating high-quality β-Ga$_2$O$_3$ with high thermal conductivity substrates such as SiC and diamond. Therefore, interface chemistry and bonding are important factors which affects interfacial thermal transport, similar to Al-diamond interfaces reported previously in the literature.[21]

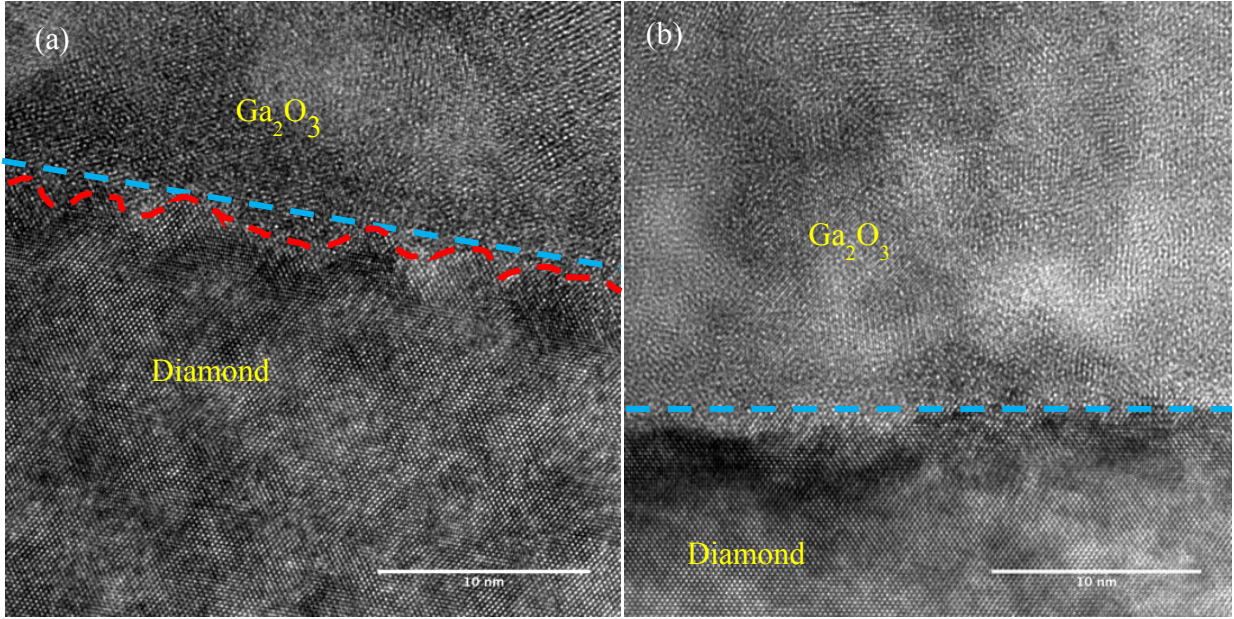

FIG. 4. TEM images of Ga$_2$O$_3$-diamond interfaces: (a) Samp2 and (b) Samp4.

**CONCLUSIONS**

Gallium oxide (Ga$_2$O$_3$) is a promising material for electronic device applications but its thermal conductivity is significantly lower than that of other wide bandgap semiconductors. Thermal management in Ga$_2$O$_3$ electronics are the key for device reliability, especially for high power and high frequency devices. This work reported the heterogeneous integration of Ga$_2$O$_3$ with diamond for thermal management of Ga$_2$O$_3$ devices. Ga$_2$O$_3$ was able to be deposited on single crystal diamond substrates by ALD. TEM studies show small grain sizes and randomly ordered grain boundaries, showing more work is required to optimize the growth conditions to yield better quality materials. The measured TBC of the $Ga_2O_3$-diamond interfaces are about 10 times larger than the Van der Waals bonded $Ga_2O_3$-diamond interfaces which facilitates thermal dissipation and indicates that interface bonding affects TBC significantly. Good contact between $Ga_2O_3$ and diamond was observed according to TEM studies. Additionally, the measured TBC of the Ga-rich and O-rich $Ga_2O_3$-diamond interfaces are about 20% smaller than that of the clean interface, indicating interface chemistry affects TBC. This study sheds light on heterogeneous integration of $Ga_2O_3$ with diamond for the benefit of thermal management of $Ga_2O_3$ devices, especially for high power and high frequency applications.

**ACKNOWLEDGEMENTS**

The authors would like to acknowledge the funding support from Air Force Office of Scientific Research under a MURI program (Grant No. FA9550-18-1-0479) and the Office of Naval Research under a MURI program (Grant No. N00014-18-1-2429). Research at the Naval Research Laboratory was supported by the Office of Naval Research.

**REFERENCES**

[1] M. Higashiwaki and G. H. Jessen, Appl. Phys. Lett. **112**, 060401 (2018).

[2] S. Pearton, J. Yang, P. H. Cary IV, F. Ren, J. Kim, M. J. Tadjer, and M. A. Mastro, Appl. Phys. Rev. **5,** 011301 (2018).

[3] P. Jiang, X. Qian, X. Li, and R. Yang, Appl. Phys. Lett. **113,** 232105 (2018).

[4] Z. Cheng, N. Tanen, C. Chang, J. Shi, J. McCandless, D. Muller, D. Jena, H. G. Xing, and S. Graham, Appl. Phys. Lett. **15** (8), (2019).

[5] Z. Cheng, L. Yates, J. Shi, M. J. Tadjer, K. D. Hobart, and S. Graham, APL Materials **7,** 031118 (2019).

[6] S. B. Reese, T. Remo, J. Green, and A. Zakutayev, Joule, **3**, 4 (2019).

[7] J. Noh, M. Si, H. Zhou, M. J. Tadjer, and D. Y. Peide, 76th Device Research Conference (DRC), pp. 1-2. IEEE, (2018).

[8] H. Zhou, K. Maize, G. Qiu, A. Shakouri, and P. D. Ye, Appl. Phys. Lett. **111,** 092102 (2017).

[9] M. J. Tadjer, J. Noh, J. C. Culbertson, S. Alajlouni, M. Si, H. Zhou, A. Shakouri, and D. Y. Peide, no. 26, pp. 1270-1270. The Electrochemical Society, (2019).

[10] D. G. Cahill, Rev. of Sci. Instrum. **75,** 5119 (2004).

[11] Z. Cheng, T. Bai, J. Shi, T. Feng, Y. Wang, M. Mecklenburg, C. Li, K. D. Hobart, T. Feygelson, and M. J. Tadjer, ACS Appl. Mater. & Interf., **11**, 20 (2019).

[12] Z. Cheng, T. Bougher, T. Bai, S. Y. Wang, C. Li, L. Yates, B. Foley, M. S. Goorsky, B. A. Cola, and F. Faili, ACS Appl. Mater. & Interf., **10**, 5, 4808 (2018).

[13] Z. Cheng, A. Weidenbach, T. Feng, M. B. Tellekamp, S. Howard, M. J. Wahila, B. Zivasatienraj, B. Foley, S. T. Pantelides, and L. F. Piper, Phys. Rev. Mater. **3,** 025002 (2019).

[14] J. T. Gaskins, G. Kotsonis, A. Giri, C. T. Shelton, E. Sachet, Z. Cheng, B. M. Foley, Z. Liu, S. Ju, and M. S. Goorsky, Nano Lett. **18**, 12, 7469, (2017).

[15] P. Jiang, X. Qian, and R. Yang, J. of Appl. Phys. **124,** 161103 (2018).

[16] F. Mu, Z. Cheng, J. Shi, S. Shin, B. Xu, J. Shiomi, S. Graham, and T. Suga, ACS Appl. Mater. & Interf., in press, (2019).

[17] D. G. Cahill, S. K. Watson, and R. O. Pohl, Phys. Rev. B **46,** 6131 (1992).

[18] M. T. Agne, R. Hanus, and G. J. Snyder, Energy & Environm. Sci. **11,** 609 (2018).

[19] A. France-Lanord, S. Merabia, T. Albaret, D. Lacroix, and K. Termentzidis, J. of Phys. Condens. Matter **26,** 355801 (2014).

[20] M. D. Losego, M. E. Grady, N. R. Sottos, D. G. Cahill, and P. V. Braun, Nat. Mater. **11,** 502 (2012).

[21] K. C. Collins, S. Chen, and G. Chen, Appl. Phys. Lett. **97,** 083102 (2010).